\documentclass{ws-wsarai}
\usepackage{hyperref}
\usepackage[super,sort&compress,comma]{natbib}
\usepackage{booktabs}
\usepackage{lipsum,pdflscape}
\usepackage{subcaption}
\usepackage{afterpage}
\usepackage{array}
\usepackage{xcolor}
\usepackage{array,ragged2e}

\definecolor{red}{rgb}{0.0, 0.0, 0.0}
\definecolor{blue}{rgb}{0.0, 0.0, 0.0}
\definecolor{yellow}{rgb}{1.0,1.0,1.0}

\newcolumntype{P}[1]{>{\RaggedRight\arraybackslash}p{#1}}

\makeatletter
\newcommand*{\@rowstyle}{}

\newcommand*{\rowstyle}[1]{%
  \gdef\@rowstyle{#1}%
  \@rowstyle\ignorespaces%
}

\newcolumntype{=}{%
  >{\gdef\@rowstyle{}}%
}

\newcolumntype{+}{%
  >{\@rowstyle}%
}
\makeatother

\begin{document}

\markboth{S. Tabakhi, M.N.I. Suvon, P. Ahadian, \& H. Lu}{Multimodal Learning for Multi-Omics: A Survey}

\catchline{}{}{}{}{}

\title{Multimodal Learning for Multi-Omics: A Survey}

\author{\bf Sina Tabakhi$^1$, Mohammod Naimul Islam Suvon$^1$, Pegah Ahadian$^2$, \and Haiping Lu$^1$\thanks{Corresponding author.}}

\address{$^1$Department of Computer Science, The University of Sheffield, Sheffield, United Kingdom}
\address{$^2$Department of Computer Science, Shahid Beheshti University, Tehran, Iran}
\emailaddress{\tt\small \{stabakhi1,mnisuvon1,h.lu\}@sheffield.ac.uk, p.ahadian@mail.sbu.ac.ir}

\maketitle

\begin{abstract}
With advanced imaging, sequencing, and profiling technologies, multiple omics data become increasingly available and hold promises for many healthcare applications such as cancer diagnosis and treatment. Multimodal learning for integrative multi-omics analysis can help researchers and practitioners gain deep insights into human diseases and improve clinical decisions. However, several challenges are hindering the development in this area, including the availability of easily accessible open-source tools. This survey aims to provide an up-to-date overview of the data challenges, fusion approaches, datasets, and software tools from several new perspectives. We identify and investigate various omics data challenges that can help us understand the field better. We categorize fusion approaches comprehensively to cover existing methods in this area. We collect existing open-source tools to facilitate their broader utilization and development. We explore a broad range of omics data modalities and a list of accessible datasets. Finally, we summarize future directions that can potentially address existing gaps and answer the pressing need to advance multimodal learning for multi-omics data analysis.\end{abstract}

\keywords{Multimodal learning; multi-omics; data fusion; machine learning; diagnostic systems; cancer prediction; open-source software.}

\section{Introduction}
Information about phenomena in the natural world typically comes from multiple modalities. Each modality from a different source represents distinct statistical properties, and multimodal data include related information from various data modalities. \textcolor{blue}{Multimodal learning presents new insights into developing machine learning models that incorporate complementary information extracted across modalities to solve complex problems \cite{ICML2011Ngiam399,baltruvsaitis2018multimodal,xu2022multimodal} in broad applications} such as healthcare data analytics,\cite{ahmed2020practicing} speech recognition,\cite{mroueh2015deep} image captioning,\cite{yu2019multimodal} human recognition,\cite{jiang2010multimodal} multimedia content retrieval,\cite{liang2018multimodal} and finance.\cite{lee2020multimodal}

Multimodal health data analytics is an important research area in multimodal learning that leverages multiple biomedical datasets to address real-world healthcare problems. Such massive datasets have been exponentially produced through technological advances in imaging, sequencing, and molecular profiling.\cite{reel2021using,kang2022roadmap} Fine-grained biological omics data can be retrieved from different high-throughput platforms, enabling scientists to explore more complex life-threatening diseases with the aim of diagnosis, prognosis, treatment, prevention, and even cure.\cite{tong2020deep,picard2021integration} \textcolor{blue}{Each omics data (i.e., genomics, epigenomics, transcriptomics, proteomics, metabolomics, microbiomics, exposomics, etc.) has demonstrated value in studying the human body’s biological processes independently.\cite{mamoshina2018machine,tabakhi2015gene,dias2020combining,arslan2021machine} Collectively, the integration of complementary information obtained from the interactions between omics data  (also referred to as multi-omics analysis) offers a promising opportunity to create a deeper understanding of most progressive diseases.\cite{acosta2022multimodal,ding2021cooperative,krassowski2020state} Designing machine learning methods for multi-omics fusion to enhance decision-making even further is still a challenging and active research area for data scientists and biologists.}

With the outstanding efforts of researchers and the vast investments of multiple institutes, notable projects have been completed and made publicly available to the research community. The Cancer Genome Atlas (TCGA)\cite{78} and The International Cancer Genome Consortium (ICGC)\cite{zhang2011international} projects have built large-scale multi-omics datasets \textcolor{blue}{with the aim of accelerating precision medicine}. The availability of this wealth of data has now revolutionized our perspective on various cancers, giving us a more meaningful understanding of entire genomic changes. This provides opportunities to achieve better performance in managing several tasks, including clinical outcome prediction, sample classification, disease subtyping, survival analysis, and biomarker identification. \cite{15_17,subramanian2020multi,picard2021integration}

\textcolor{blue}{Although researchers have access to a wealth of various omics data, extracting information from them for clinical prediction is complex and remains a challenge.} Multi-omics data involve different high-dimensional data with a relatively small number of patients, known as `the curse of dimensionality', which gives rise to the overfitting of models \textcolor{blue}{with the lack of generalizability}. \cite{acosta2022multimodal, momeni2020survey} Moreover, the utilization of different platforms to generate omics data presents heterogeneous datasets with varying data distributions and types that need to be adequately tackled. \cite{kang2022roadmap,picard2021integration} Another problem is related to \textcolor{blue}{measurement errors} in sequencing technologies that produce missing values in omics datasets. \cite{mirza2019machine,kang2022roadmap} Complexity and noisiness lead to other challenges that are the nature of biological omics data and make multi-omics analysis difficult to work on. \cite{momeni2020survey,picard2021integration} Furthermore, multi-omics datasets are commonly imbalanced at the levels of class and/or feature, which biases learning models towards majority classes or omics with larger numbers of features. \cite{mirza2019machine,reel2021using} The details of these challenges and their related sub-challenges are provided in Section \ref{sec:challenges}.

\afterpage{%
\begin{landscape}
\begin{table}[!t]
\vspace*{-1cm}
  \caption{A summary of survey papers on multi-omics analysis published since 2020.}
  \renewcommand{\arraystretch}{1.4}
  \label{table:survey}
  \centering
    \hspace*{-2cm}
    \makebox[\textwidth]{\resizebox{1.75\textwidth}{!}{\begin{tabular}{p{0.17\linewidth}p{0.04\linewidth}p{0.22\linewidth}p{0.23\linewidth}P{0.22\linewidth}P{0.17\linewidth}P{0.17\linewidth}P{0.17\linewidth}}
    \toprule
    Reference & Year & Scope & Challenges & Fusion approaches & Datasets & Platforms & Tools\\
    \midrule
    \citet{kang2022roadmap} & 2022 & A comprehensive overview of multi-omics data integration using deep learning & 
    The curse of dimensionality \newline
    Missing values \newline
    Heterogeneity
    & 
    Neural network-based
    & N/A & N/A & N/A\\
    \citet{reel2021using} & 2021 & A background in multi-omics data with an exploration of integration strategies and key challenges & 
    The curse of dimensionality \newline
    Class imbalance \newline
    Heterogeneity 
    & 
	Early fusion \newline
	Late fusion  \newline
	Transformation  
    & N/A & N/A & N/A\\
    \citet{picard2021integration} & 2021 & A mini-review of different challenges and integration approaches for multi-omics data analysis & 
    The curse of dimensionality \newline
    Class and feature imbalance \newline
    Missing values \newline
    Heterogeneity 
    & 
    Early fusion \newline
    Late fusion \newline
    Mixed fusion \newline
    Hierarchical fusion \newline
    Factorization 
    & N/A & N/A & N/A\\
    \citet{15_17} & 2021 & A theoretical and practical review of several representative joint dimensionality reduction methods & 
    The curse of dimensionality \newline
    Heterogeneity 
    & 
    Graph-based \newline
    Probabilistic model-based \newline
    Factorization 
    & N/A & N/A & A single tool to implement several factorization-based methods \\
    \citet{subramanian2020multi} & 2020 & A review of the tools and methods for multi-omics integration and the introduction of visualization portals & 
    The curse of dimensionality \newline
    Heterogeneity 
    & 
    Kernel-based \newline
    Graph-based \newline
    Probabilistic model-based \newline
    Matrix-based factorization
    & A few public multi-omics datasets & A set of portals for visualization and analysis of multi-omics data & A range of tools, each implemented for a specific multi-omics integration method\\
    \citet{momeni2020survey} & 2020 & A complete overview of the multi-omics challenges with a focus on feature selection and data fusion methods & 
    The curse of dimensionality \newline
    Class imbalance \newline
    Missing values \newline
    Heterogeneity \newline
    Complexity and noisiness 
    & 
    Early fusion \newline
    Late fusion \newline
    Transformation 
    & Several cancer datasets in the TCGA & N/A & N/A\\
    \citet{nicora2020integrated} & 2020 & An overview of multi-omics integration methods developed in oncology & 
    The curse of dimensionality 
    &
    Kernel-based \newline
    Graph-based \newline
    Neural network-based \newline
    Probabilistic model-based \newline
    Factorization 
    & N/A & N/A & A complete list of tools published in oncology papers\\
    \citet{nguyen2020multiview} & 2020 & A review of multiview learning methods in multi-omics analysis and their applications to biological systems & 
    The curse of dimensionality \newline
    Heterogeneity \newline
    Noisiness 
    & 
    Transformation \newline
    Factorization 
    & Some multi-omics benchmark datasets  & N/A & N/A\\
\botrule
  \end{tabular}}}
\end{table}
\end{landscape}
\clearpage   %
}

Several efforts have been made to review multi-omics analysis. A summary of survey papers published on multi-omics analysis since 2020 is provided in Table \ref{table:survey}. \textcolor{blue}{In terms of data challenges, most surveys have reviewed several of them but lack a comprehensive and structured list, and some challenges have yet to be discussed in detail.} On the other hand, each paper has proposed different ways to categorize the current fusion methods. Although existing methods have been reviewed under these categorizations, a new class of techniques, joint feature selection, has not been explored. Moreover, some publications have used a flat organization for integration approaches, which does not reflect the similarities and differences among methods. Beyond these issues, public datasets and platforms are not well-studied in existing surveys. \textcolor{blue}{There is an increasing need to track high-quality, real-world benchmarks for reproducible research that can be achieved via data sharing.} Furthermore, open-source software with the implementation of key methods is an essential resource that supports researchers in building their frameworks rapidly and having a comprehensive comparison in their experimental evaluation. Therefore, highly reusable software can accelerate multi-omics research progress.

\textcolor{blue}{In this survey, we provide a comprehensive overview of multimodal learning for multi-omics data, specifically focusing on the available data challenges, fusion approaches, and open data and tools.} There are four contributions.

\begin{romanlist}[(ii)]
 \item We summarize a broad range of omics types, including radiomics and phenomics, their applications, and their relationships, as well as frequently used multi-omics datasets and online portals to discover datasets.
 \item We categorize key data challenges comprehensively with well-structured sub-challenges to offer new perspectives on their similarities and differences. 
 \item We group multi-omics fusion approaches into a new multi-level organization with a concise review of these approaches, particularly introducing a finer sub-categorization for transformation and factorization and a new category termed joint feature selection.
 \item We collect active open-source software packages based on several criteria, including recency, citation, and GitHub popularity. To the best of our knowledge, this is a brand new collection with no similar ones in the literature of multi-omics analysis. This list can help researchers become more aware of available resources and contribute to the multi-omics research community effectively and efficiently.
\end{romanlist}

The rest of this survey is organized as follows. Section \ref{sec:omics-data} presents a brief description of omics data. Section \ref{sec:challenges} investigates challenges and related sub-challenges of multi-omics data analysis. Section \ref{sec:fusion} explores various fusion approaches in multi-omics integration. Section \ref{sec:tools} introduces open-source software available for multi-omics analysis. Section \ref{sec:dataset} reviews open-access multi-omics datasets along with portals for free download. Section \ref{sec:discussion} discusses existing gaps and promising future research directions. Finally, Section \ref{sec:conclusion} \textcolor{blue}{concludes} this study.

\section{Multi-Omics Data}
\label{sec:omics-data}
\begin{figure}[!t]
    \centering
    \includegraphics[width=13cm]{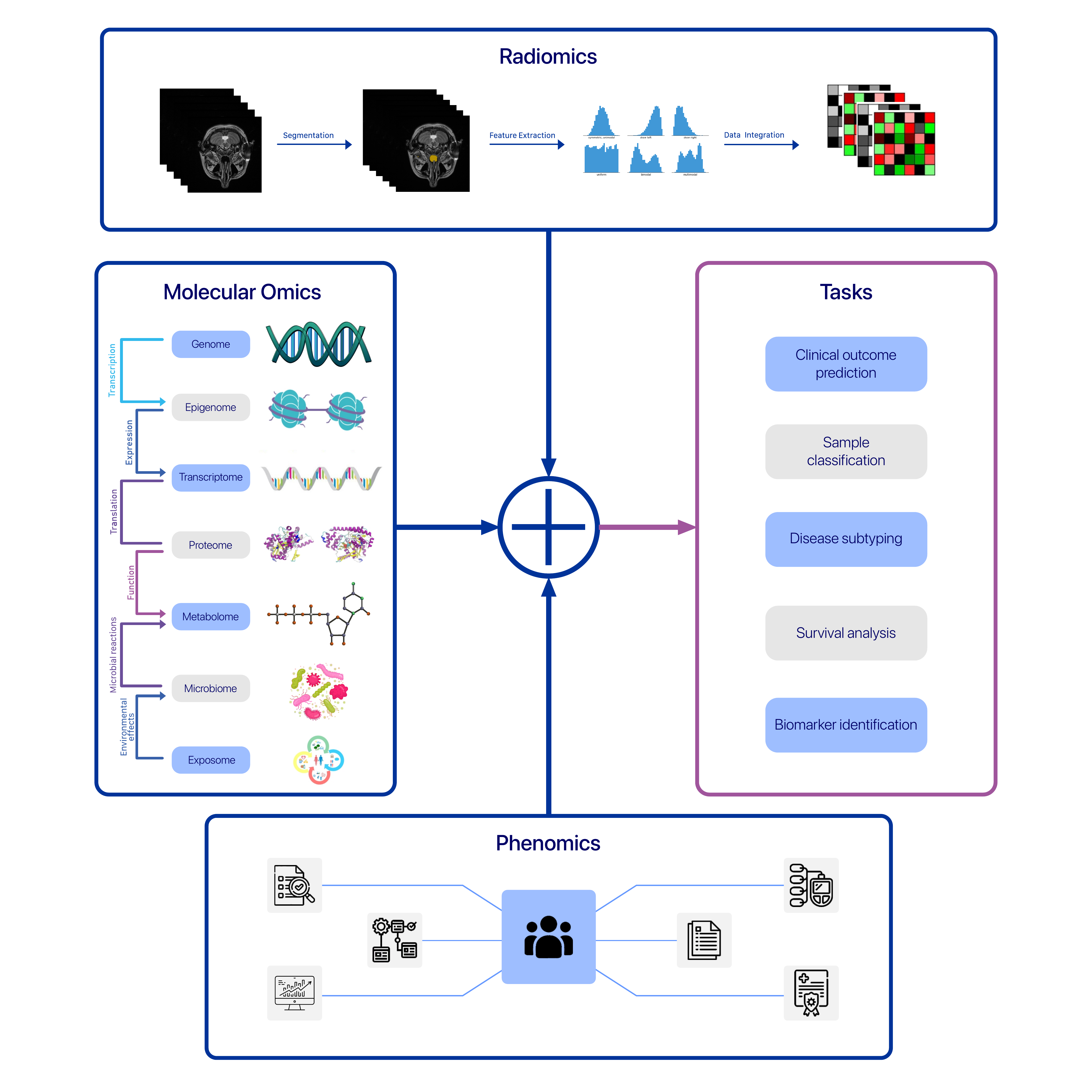}
    \caption{ Multiple omics modalities and their fusion for various tasks.}
    \label{fig:omics-pipeline}
\end{figure}

Multi-omics data are derived from different sources: molecular omics data, radiomics, and phenomics. Molecular omics data are the traditional dimension of multi-omics data, which aim to analyze the molecular biology of human diseases and consist of genomics, transcriptomics, proteomics, metabolomics, epigenomics, \textcolor{red}{microbiomics, and exposomics}. Radiomics, on the other hand, studies medical images collected from medical imaging technologies.\cite{lambin2012radiomics} Phenomics is another valuable dimension of multi-omics data that comprises clinical and biochemical data of individuals. \cite{bilder2009phenomics} Each source of omics data has been separately studied to answer biomedical questions. While different sources of information are complementary and contribute to the discovery of unique aspects of diseases, the fusion of multiple omics modalities has widely been done in existing research. Therefore, introducing each modality of omics data and their potential applications in solving biomedical tasks helps researchers consider all omics modalities in the long run.\cite{karczewski2018integrative,acosta2022multimodal} Table \ref{table:omics} provides an overview of multiple modalities of omics data and their applications applied to biomedical studies. Besides, the multiple omics data and their fusion for various tasks are visually presented in Figure \ref{fig:omics-pipeline}.

We review multi-omics data in the following two subsections based on their prevalence in the research field, including traditional molecular omics data and non-molecular omics data.

\afterpage{%
\begin{landscape}
\begin{table}[!t]
    \vspace*{-1.5cm}
  \caption{Overview of multi-omics data and their various applications including diagnosis, prognosis, and precision medicine.}
  \renewcommand{\arraystretch}{1.4}
  \label{table:omics}
  \centering
    \hspace*{-3cm}
    \makebox[\textwidth]{\resizebox{1.6\textwidth}{!}{\begin{tabular}{=p{0.13\linewidth}+p{0.65\linewidth}+p{0.23\linewidth}+p{0.42\linewidth}}
    \toprule
    Omics name & Description & Types/Technologies & Applications
    \\
    \midrule
    Genomics & 
    The exact structure of a DNA molecule can be determined in genomics to learn more about a patient's molecular biology and genetic variants associated with disease, response to treatment, or future patient prognosis. & 
    Structural genomics\cite{59}\newline
    Functional genomics\cite{60} \newline
    Comparative genomics\cite{61} \newline
    Mutation genomics\cite{62} 
    &
    Gene discovery and diagnosis of rare monogenic disorders (e.g., heart disease and cancers) \newline
    Identification and diagnosis of genetic factors contributing to prevalent diseases \newline
    Pharmacogenetics and targeted therapy \newline
    Diagnosis of infectious diseases \\
    
    Epigenomics & 
    The organization and regulation of DNAs in cells are explored by epigenomics to promote a stably heritable phenotype without changes in the DNA sequence. This exploration is essential for controlling normal development and homeostasis. & 
    DNA methylation\cite{63} \newline
    Histone modification\cite{64}  \newline
    Non-coding RNA\cite{65}
    &
    Early-stage detection, e.g., coronary artery disease \newline
    Therapeutic development \\
    
    Transcriptomics & 
    The transcriptome, a collection of all the gene readouts in a cell, is investigated in transcriptomics to identify the functions of genes. & 
    DNA microarrays\cite{66}  \newline
    RNA-Seq\cite{67}
    &
    Identification of early cancer biomarkers \newline
    Human and pathogen transcriptomes\newline
    Response to the environment\newline
    Gene function annotation\newline
    Non-coding RNA\\
    
    Proteomics & 
    The use of technology to identify and quantify the entire protein content of a cell, tissue, or organism can be found in proteomics. & 
    Expression proteomics\cite{68}  \newline
    Functional proteomics\cite{69}  \newline
    Structural proteomics\cite{70}   
    &
    Protein production, degradation, and steady-state abundance rates \newline
    Protein movement between subcellular compartments\newline
    Protein involvement in metabolic pathways\newline
    Protein interaction display used in the drug discovery process\\
    
    Metabolomics & 
    The substrates and products of metabolism influenced by genetic and environmental factors are studied in metabolomics. & 
    Metabolite fingerprinting\cite{71}  \newline
    Metabolic profiling \newline
    Targeted analysis
    &
    Investigation of several human diseases (e.g., cancers and natural metabolic errors) \newline
    Design improved therapeutic strategies \newline
    Toxicology (i.e., toxicological effects) \newline
    Pharmacology (i.e., nutrition)\\
    
    \rowstyle{\color{red}}
    Microbiomics &  
    The analysis of molecules involved in the structure and function of microbial communities in the human body is investigated in microbiomics to diagnose human diseases. & 
    Microbiome \cite{schwabe2013microbiome} &
    Infectious disease diagnosis \newline
    Microbial components monitoring for noncommunicable chronic diseases (e.g., heart disease, diabetes, and chronic lung disease)\\
    
    \rowstyle{\color{red}}
    Exposomics &  
    Understanding how environmental exposures affect the unique human characteristics to develop diseases is explored in exposomics. &
    Environmental exposures \cite{price2022merging} 
    &
    New insights into the development of chronic diseases\newline
    Reveal nongenetic disease causes
    \\
    
    Radiomics & 
    The extraction of quantitative features from medical images through mathematical algorithms is described in radiomics to reveal patterns and characteristics of various cancers. & 
    Medical images \newline
    &
    Diagnostic differentiation of suspected tissue \newline
    Survival prognosis \newline
    Prediction of clinical responses\newline
    Prediction risk of distant metastasis\cite{72}\\
     
    Phenomics & 
    The understanding of variations in phenotypic characteristics of humans as a result of interactions between the environment and genotypic is covered in phenomics. It advances biomedical research and personalized medicine. & 
    Qualitative  traits\cite{lanktree2010translational}  \newline
    Quantitative traits
    &
    Functional genomics \newline
    Pharmaceutical research\newline
    Agricultural research\newline
    Phylogenetics \\
\botrule
  \end{tabular}}}
\end{table}
\end{landscape}
\clearpage   %
}

\subsection{Traditional molecular omics data}
The human body is made of billions of building blocks called cells, and genes are placed inside them. Genes are small parts of deoxyribonucleic acid (DNA) with complex structures that produce proteins for the development and maintenance of the human body. As a result, studying various constituents of genes helps diagnose, treat, and cure life-threatening human diseases such as cancers.

The rapid development of DNA sequencing into next-generation sequencing (NGS) has generated diverse molecular data named omics. \cite{karczewski2018integrative, 96} It has enabled bioinformaticians to analyze molecular omics, which offers a biological strategy for combining genomics with transcriptomics, proteomics, metabolomics, and epigenomics. Consequently, this integration provides knowledge about gene expression, gene activation, and protein levels, which play a vital role in responding to several biological tasks, including disease diagnosis, prognosis, treatment, prevention, and biomarker identification.

The complete set of genetic information for human development and growth is found in the genome and expressed by ribonucleic acid (RNA). The main aim of \textbf{genomics} is the study of the genome for extracting information from genes to determine variations in DNA sequences that can give rise to human diseases. The influence of diseases on the function of cells or living organisms is examined by \textbf{transcriptomics} through gene expression. In other words, transcriptomics identifies modifications in RNA, which is a decisive factor in the occurrence of diseases. \textbf{Proteomics}, on the other side, investigates the set of proteins produced in an organism, system, or biological setting to complement gained information from genomics. This type of omics data is generated by a key analytical technology called mass spectrometry (MS).\cite{95} \textbf{Metabolomics} presents a fingerprint of the physiology of the cells by studying chemical processes involving cellular metabolites and their interactions to detect metabolic diseases and discover biomarkers. \textbf{Epigenomics} examines the modifications in gene function regulation without altering the DNA sequences. The availability of epigenomics can highlight a way to integrate gene expression changes and environmental cues that reveal complementary information in human disease detection. \textcolor{red}{Another factor in shaping human diseases is microbiomes which consist of bacteria, archaea, fungi, algae, and protists. \textbf{Microbiomics} or microbiome science is an emerging area that investigates communities of particular microbial, known as microbiota, to identify microbes and viruses that result in disease. Beyond genetic, environmental exposures also play a decisive factor in several human diseases, such as skin disorders. \textbf{Exposomics} is the study of measuring the total environmental exposures during the individual’s life to explore the harmful effects of exposures on health and understand the complicated interaction between genetic and environmental factors.}

Several other omics types have been discovered but have not yet been incorporated into molecular omics analysis. \textbf{Metagenomics}, \textbf{metatranscriptomics}, and \textbf{metaproteomics} are additional molecular omics for which bioinformatics still encounters critical obstacles in their application, such as short reads of sequencing technologies, a high error rate in sequencing technologies, and economic reasons.\cite{106} Furthermore, \textbf{glycomics} is a complete snapshot of glycans constructed in the cells, and its link with other types of omics data has yet to be developed. \cite{74}

\subsection{Non-molecular omics data}
With the progress in medical imaging technologies and the discovery of the importance of clinical data, complementary sources of omics data, called radiomics and phenomics, have emerged to add meaningful knowledge in solving biological problems.

In precision medicine, \textbf{radiomics} aims to extract quantitative features from medical images and analyze them to enable decision-making systems for a more accurate diagnosis.\cite{104} The workflow of radiomics starts with acquiring image data generated in different modalities, such as computed tomography (CT), magnetic resonance imaging (MRI), and digital histopathology. When images have been collected, segmentation is typically carried out to find essential regions of images to simplify their representation. After that, the feature extraction process is executed to produce different feature types, such as histograms and shape-based features. The valuable features are then combined into specific matrices through mathematical algorithms. Finally, the outcome assessment is made based on machine learning models.\cite{caruso2021radiomics} A critical step in the described workflow is access to the standardized image datasets. One of the big projects in collecting and organizing medical images of cancer is The Cancer Imaging Archive (TCIA), which facilitates easy and public access to images related to common cancers (e.g., lung cancer). \cite{clark2013cancer}

On the other hand, \textbf{phenomics} systematically measures the full set of qualitative and quantitative phenomes, including physical and biochemical traits. Phenomics examines the effect of an individual’s environment and lifestyle on their genome-wide scale to assess the risk of diseases.\cite{houle2010phenomics} Phenomics includes electronic health record (EHR)-based phenotyping, the study of extracting information from EHRs to discover clinical characteristics of individuals that can improve our understanding of human health.\cite{richesson2013electronic} The genotype-phenotype relationship is an integrative point of view of genetic and phenotypic levels, which causes a unique understanding of complex human diseases. \cite{bilder2009phenomics}

\begin{figure}[!t]
    \centerline{\includegraphics[trim=45 650 0 30, width=7.9in]{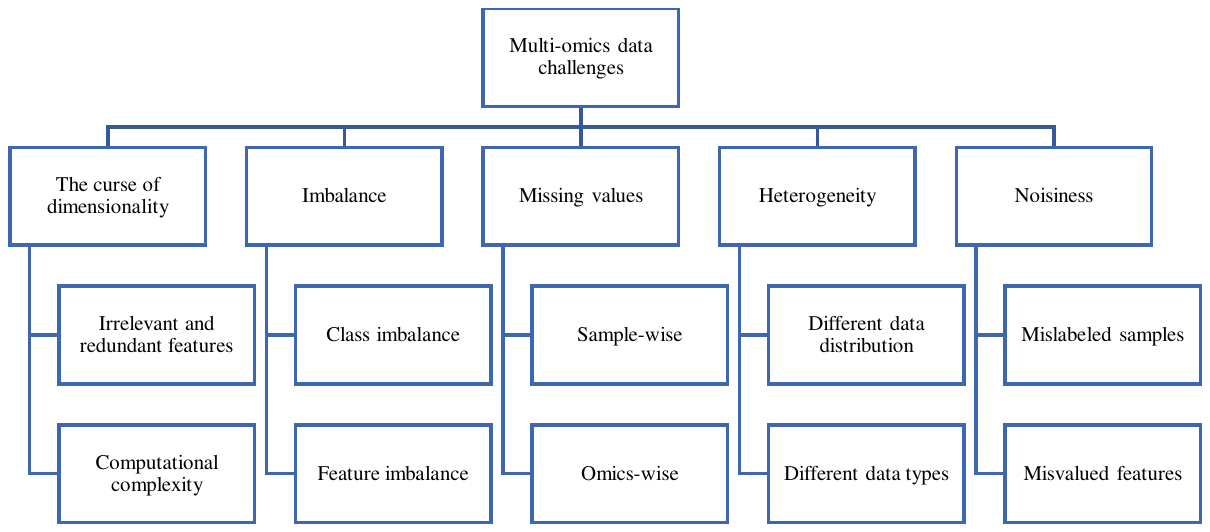}}
    \caption{Challenges and related sub-challenges in multi-omics data.}
    \label{fig:challenges}
\end{figure}

\section{Challenges in Multi-Omics Data Analysis}
\label{sec:challenges}
Working with multi-omics data has always resulted in a broad range of challenges that adversely affect learning tasks and should be overcome to improve performance. The arising challenges are fundamental to machine learning research and broadly fall into five categories briefly introduced in this section. Figure \ref{fig:challenges} summarizes the challenges and related sub-challenges in multi-omics data analysis.

\subsection{The curse of dimensionality}
\label{sec:curse-dimension}
Biological data usually suffer from the small sample size problem because data collection is financially costly, and the number of clinical research participants is limited.\cite{acosta2022multimodal, picard2021integration} Therefore, most types of omics datasets contain a large number of features compared to only a relatively small number of patients, leading to the classical phenomenon in machine learning named the curse of dimensionality.\cite{de2018computational,15_17} Although many features are available in omics data, their correlation is very high, and some features do not highlight disease-specific indicators.\cite{hira2015review,kirpich2018variable} It becomes even more challenging when irrelevant and redundant features in each omics data type are integrated into the multi-omics analysis. Therefore, these features mislead the learning algorithm and limit the model's generalizability to unseen samples. Moreover, the computational complexity of developing a model will be substantially increased in the presence of high-dimensional datasets. \cite{mirza2019machine,hira2015review,tabakhi2015gene}

To tackle these problems, dimensionality reduction techniques have been applied to multi-omics data in two ways: feature selection and feature extraction.\cite{wang2016feature,picard2021integration} Feature selection refers to methods that select a small number of features from the original set by removing irrelevant and redundant features.\cite{wu2019selective} Many studies have independently utilized feature selection on each omics data as a pre-processing step.\cite{chen2020integrative,wang2021mogonet,zhang2019integrated,torres2022human,lualdi2019statistical,goh2016evaluating,smit2008statistical} However, the redundancy among features across omics datasets can be missed, resulting in low-performance multi-omics analysis. A few researchers have recently tried to use feature selection by considering all omics datasets together to capture the interaction between different omics types.\cite{el2018min,tabakhi2022magentomics} Despite the consideration of molecular layer interactions, the computational cost of these methods is high. 

Another strategy to reduce the dimensions of the input data is feature extraction, in which a new representation is constructed by projecting the initial feature set into a new space with lower dimensions. \cite{meng2016dimension} The two frequently used methods are principal component analysis (PCA)\cite{ringner2008principal} and its extensions (i.e., kernel PCA, \cite{scholkopf1998nonlinear,feng2021multi} Bayesian PCA, \cite{nounou2002bayesian} sparse PCA, \cite{shen2008sparse} and consensus PCA \cite{meng2016dimension}) and linear discriminant analysis (LDA).\cite{martinez2001pca} Other methods proposed based on this approach are partial least squares, \cite{bersanelli2016methods} canonical correlation analysis, \cite{witten2009penalized} independent component analysis,\cite{sompairac2019independent} and multiblock discriminant analysis.\cite{kang2015multiblock} 
\textcolor{red}{Graph representation learning is another promising technique for overcoming the curse of dimensionality by leveraging the graph structure, which has been investigated in several studies.\cite{hamilton2020graph,amor2021graph,zeng2019deepdr,xu2023gripnet,li2022graph,Ektefaie2021machine}}

\subsection{Imbalance problem}
Another frequent problem with multi-omics data is the imbalance problem, which can be seen at the class or feature level. Class imbalance naturally occurs in multi-omics data when disease-specific classes are rare and comprise a small proportion of samples compared to other classes. This imbalance gives rise to the machine learning model’s bias toward the majority classes, and the danger of overfitting increases.\cite{roelofs2019meta,li2010learning} Over and under-sampling \cite{lin2016imbalanced,wang2022large,johnson2019survey} and cost-sensitive learning \cite{zhou2005training} are the most widely used machine learning techniques for balancing datasets.

Feature imbalance, on the other hand, refers to the different distribution of feature dimensionality within each omics data. One omics may have only hundreds of features, while another may contain thousands. As a result, the learning model is more likely to pay more attention to omics with a larger number of features, leading to a learning imbalance.\cite{picard2021integration,hira2021integrated} To deal with this problem, various strategies have been introduced, ranging from dimensionality reduction (see Section \ref{sec:curse-dimension} for details) to late fusion (see Section \ref{sec:late-fusion} for details).

\subsection{Missing value}
Missing values are an inevitable issue with multi-omics data that can occur in the process of omics data acquisition via high-throughput platforms. \textcolor{red}{Data missingness can adversely affect the machine learning algorithms, leading to biased results and poor induction models.\cite{saar2007handling,emmanuel2021survey}} Two forms of missing values include sample-wise and omics-wise missingness.\cite{song2020review,kang2022roadmap} Sample-wise missing values are a traditional problem in machine learning, where part/all of the samples have missing values. \textcolor{red}{While the simplest solution to this problem is to exclude patients with missing values,\cite{el2018min} the risk of losing significant information from the datasets increases and may result in selection bias.\cite{haneuse2021assessing,acosta2022multimodal}} These gaps can be filled with missing value imputation, a pervasive statistical strategy for replacing missing values with statistical estimates. Several imputation methods have been introduced in the literature, such as mean, median, and $k$-nearest neighbors.\cite{theodoridis2006pattern,chen2019sparse,feng2021addressing} Beyond single omics imputation, there is an increasing interest in integrative imputation strategies for achieving better results. \cite{lin2016integrative,barbeira2018exploring,zhang2019integrative,song2020review,voillet2016handling}

In omics-wise missingness, biological samples are represented in some, but not all, omics datasets, which are referred to as `partial datasets.' This issue exists in multi-omics data because of economic reasons, technical constraints, or experimental limitations.\cite{kang2022roadmap} Since most models in machine learning require complete datasets, a naive way to resolve this issue is to remove patients with missing omics. However, the data collection cost is high, and it is more appropriate to consider all available samples to address partial datasets.\cite{fang2018bayesian} Several researchers have developed models to work directly on partial datasets.\cite{rappoport2019nemo,li2014partial,xu2021network,rappoport2020monet}

\subsection{Heterogeneity}
Each omics type has been collected using different high-throughput technologies, which have produced heterogeneous datasets. The heterogeneity of multi-omics data generates various data distributions and different data types (e.g., discrete, continuous, numerical, and categorical).\cite{mirza2019machine} This variety of data distributions and types causes a considerable challenge to the learning model. Therefore, utilizing appropriate normalization and scaling approaches before any analysis has an essential role in achieving better performance.\cite{krassowski2020state,tarazona2020harmonization,Akdemir857425} Another way to deal with heterogeneity is using late fusion to build a different model for each omics data and then combine the results (see Section \ref{sec:late-fusion} for details).

\subsection{Noisiness}
Complex workflows for generating large-scale omics data have produced noisy multi-omics data, which can arise as mislabeled samples or misvalued features.\cite{yoo2021community, zych2017regenotyper} Sample mislabeling is a well-known problem in healthcare research, and these noisy datasets can misguide the learning algorithms to find misleading patterns. Therefore, some necessary steps should be taken to rectify the errors. Mislabeled sample correction by matching methods has been used in several research works.\cite{yoo2021community,huang2013tool,lee2019probabilistic,javed2020detecting} Since there are unreliable samples in multi-omics data, unsupervised and self-supervised methods may perform better than supervised methods, as reviewed in some papers.\cite{nguyen2019pinsplus,tini2019multi,tabakhi2022magentomics,vahabi2022unsupervised}

Misvalued features, on the other hand, can occur due to measurement errors or biological deviations.\cite{picard2021integration} Consequently, extracting meaningful information from multi-omics data is a challenging task that needs to be addressed appropriately. One technique to address this problem uses regulatory network strategies (e.g., RNA-protein interaction networks) to reduce noise based on interactions between biological datasets.\cite{ma2019integrate} Some other efforts have been made to solve this problem in the literature.\cite{zhang2022integrative,duan2021evaluation}

\section{Multi-Omics Fusion Strategies}
\label{sec:fusion}
\begin{figure}[!t]
    \centerline{\includegraphics[trim=60 610 0 20, width=8in]{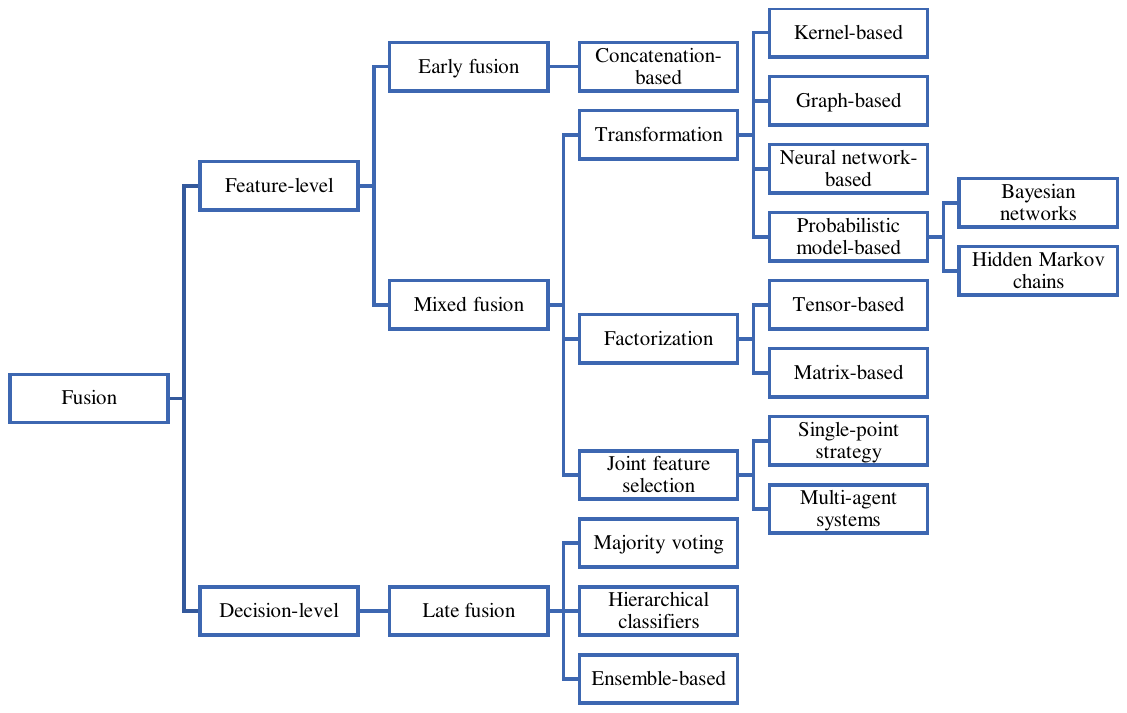}}
    \vspace*{8pt}
    \caption{Categorization of multi-omics data fusion approaches.}
    \label{fig:fusion}
\end{figure}

Multi-omics data fusion integrates information from multiple omics modalities to capture intra-omics and cross-omics interactions for a deeper insight into biological processes and better decision-making. It can leverage the interactions and supplementary information among omics data to provide richer knowledge compared to single omics learning.\cite{103} Researchers have proposed a growing number of methods to fuse multi-omics data. We categorize all developed methods as \textcolor{blue}{shown in Figure \ref{fig:fusion}} and discuss them in the following subsections. 

\subsection{Feature-level fusion}
Feature-level fusion aims to integrate extracted features from input multi-omics data to capture richer information. Then, a learning model is applied to the integrated features to carry out downstream tasks.\cite{singh2019comprehensive,Ross2009fusion} Generally, feature-level fusion comprises early fusion, transformation, factorization, and joint feature selection strategies described in the following sections.

\subsubsection{Early fusion}
The early fusion approach directly concatenates each omics dataset to construct a single extensive dataset containing all features before being fed to learning algorithms.\cite{adossa2021computational} Methods based on this approach benefit from cross-modality learning, which refers to learning by involving information obtained from multiple modalities. The development of this strategy is simple, and the final joint dataset can be used as the input for numerous classical machine learning algorithms such as artificial neural networks (ANNs),\cite{kim2013athena} support vector machines (SVMs),\cite{ma2016breast} decision trees (DTs),\cite{lin2017machine} random forests (RFs),\cite{ma2020diagnostic} and $k$-nearest neighbors ($k$-NN).\cite{wang2020moronet} Training only one single algorithm in early fusion leads to a more straightforward pipeline for implementing this approach. \cite{baltruvsaitis2018multimodal}

Despite the simplicity of early fusion, the newly generated single dataset has a higher dimensionality for a relatively smaller number of patients. This problem makes the model’s training difficult, decreases the performance, and increases the computational time.\cite{ritchie2015methods} Since there is a noticeable difference in feature dimensionality in multi-omics data, another drawback of early fusion is the model's tendency to learn more from the omics with a larger number of features.\cite{cavill2016transcriptomic} Employing dimensionality reduction techniques as a pre-processing step can reduce the adverse effects of these challenges by keeping a small number of discriminating features \cite{spicker2008integration,worheide2021multi,mirza2019machine, wang2016integrative, 15_17} (see Section \ref{sec:curse-dimension} for details).

\subsubsection{Mixed fusion}
\label{sec:fusion-mix}
In the mixed fusion approach, each original omics dataset is separately transformed into a new intermediate representation, after which they are merged to produce the final representation. At this stage, machine learning algorithms can be employed in the joint representation of multi-omics data.\cite{picard2021integration} Since the intermediate representations have a lower dimensionality, future analysis can be done more efficiently. Furthermore, using independent representations in the first step makes it possible to address the heterogeneity of multiple modalities.\cite{ritchie2015methods} Transformation, factorization, and joint feature selection are three categories of mixed fusion to be discussed in the following subsections. 

\paragraph{Transformation.}
The goal of transformation-based methods is to project each unimodal dataset into a new subspace and combine these generated subspaces before building a learning model. These methods can integrate a different range of omics modalities for diagnostic tasks.\cite{lin2017machine} Strategies for generating the new subspace can be divided into four categories, kernel-based, graph-based, neural network-based, and probabilistic model-based fusions.

\subparagraph{Kernel-based fusion} is a well-known form of transformation that uses kernel functions to map original features onto a new space with higher dimensions. Kernels allow such methods to work in high-dimensional space to explore similarities and relationships between samples.\cite{yan2017comparison} Several widely used kernel functions are linear, polynomial, sigmoid, hyperbolic tangent, string, tree, graph, Gaussian, and radial basis functions.\cite{roman2021depth} Kernel-based methods are capable of applying different kernels to multiple omics datasets that provide various similarity metrics. SVM is a traditional machine learning algorithm for working with kernels.\cite{cortes1995support,noble2004support,ben2008support} Multiple kernel learning (MKL) is another algorithm that utilizes different kernels for multiple omics modalities to find correlations across modalities and consolidates them into a single kernel for further analysis. \cite{gonen2011multiple,baltruvsaitis2018multimodal,lanckriet2004statistical,seoane2014pathway} Other kernel-based fusion methods include semi-definite programming (SDP), SDP/SVM,\cite{lanckriet2004statistical} sequential minimal optimization MKL (SMO-MKL)\cite{tao2019classifying}, relevance vector machine (RVM)\cite{tipping2001sparse}, Ada-boost RVM\cite{wu2010prediction}, and kernel PCA.\cite{mariette2018unsupervised}

Despite having attractive performance, this fusion approach is computationally expensive compared to other transformation-based techniques.\cite{reel2021using}

\subparagraph {Graph-based fusion} \textcolor{red}{is becoming a prevalent technique for integrating multimodal data in biomedical and healthcare studies due to its capacity to capture molecular interactions.\cite{li2022graph}} This type of fusion is broadly accomplished in two ways. The first way models each modality as a graph and combines them to make a unified graph for establishing further analysis.\cite{chierici2020integrative} In the graphs created for each omics, nodes represent samples, and edges represent relationships between pairs of samples. These graphs are subsequently converted to similarity matrices in an iterative optimization process\cite{wang2014similarity} or a single iteration algorithm.\cite{rappoport2019nemo} In the final stage, matrices are fused to construct a single graph that can be fed into a machine learning model for performing learning tasks.\cite{picard2021integration} Similarity network fusion (SNF) is a graph-based method that fuses constructed similarity graphs of patients through an iterative process of updating similarities to identify cancer subtypes derived from a clustering algorithm.\cite{wang2014similarity} Ranked SNF is an extension of SNF that uses a feature-ranking strategy for computing multi-omics features’ rankings to build a final graph before applying spectral clustering.\cite{chierici2020integrative} \citet{rappoport2019nemo} proposed a three-stage graph-based fusion algorithm for clustering called NEighborhood-based Multi-Omics (NEMO) clustering. In the first stage, a similarity matrix was created for each omics modality based on patient relationships. Then, matrices were fused to generate a relative similarity matrix in a single iteration. Finally, the spectral clustering algorithm was used to cluster cancer samples. \citet{ramirez2020classification}, \citet{kim2015knowledge}, and \citet{wen2021multi} have presented more fusion methods based on graph.

\textcolor{red}{In contrast to the previous strategy that fuses unimodal graphs directly, graph representation learning (also known as graph embedding) integrates latent representation spaces of graphs into a joint representation and feeds into machine learning models. In other words, each graph-structured input modality is encoded into a low-dimensional space to reflect the graph topology while preserving the original graph structure. Then, latent representations are combined to perform the downstream task.\cite{li2022graph,Ektefaie2021machine} \citet{amor2021graph} investigated multimodal learning on tissue-specific multi-omics data using a graph embedding model based on the variational autoencoder. In the first phase, RNA sequencing and gene methylation datasets were independently transformed into compact vectorial spaces using two graph convolutional neural networks. These representations were then incorporated and fed into the variational graph autoencoder model for the purpose of link prediction. \citet{zeng2019deepdr} proposed a graph-based fusion architecture using deep learning (deepDR) for in silico repositioning of drugs. They applied the random walk approach to convert each drug's structure into a vector representation, which they subsequently fused via a multimodal deep autoencoder for a prediction task. Graph information propagation network (GripNet) is a general framework to integrate several modalities using heterogeneous graph representation learning. In this framework, a new data structure named supergraph was defined to embed each modality in a compact space and pass messages between them for performing a specific task. \cite{xu2023gripnet}}

Since graphs are formed based on samples rather than features, the complexity of the whole pipeline does not significantly increase by adding new omics modalities. 

\subparagraph{Neural network-based fusion} is a growing approach in the multi-omics research area due to its superior performance in numerous domains of multimodal learning.\cite{ramachandram2017deep} In this approach, a network is trained with each modality from biological systems to learn a joint representation of the inputs. The hidden layers of the built networks are then passed into another neural network for more analysis.\cite{mroueh2015deep} The benefits of using neural networks for fusion are hierarchical representations via layers of neural networks, learning complex non-linear relationships of features, and scalability in terms of the number of omics.\cite{kang2022roadmap} \citet{bica2018multi} proposed a new neural network approach for the fusion of multi-omics data derived from TCGA. They used two feed-forward neural networks, each receiving specific omics data, to obtain cross-correlations in multiple omics data and combine them into a fully connected network for the prediction tasks. \citet{lee2019mildint}  and \citet{alkhateeb2021deep} presented other  fusion methods based on neural networks.

An unsupervised class of neural network-based fusion is autoencoders that learn compact representations of input omics data through the encoder-decoder structure. \citet{poirion2018deep} introduced an algorithm in which an autoencoder is built for each omics data to link them for inferring survival subtypes. \citet{zhang2018deep} presented an unsupervised multi-omics integration method based on an autoencoder with three hidden layers to identify prognosis subtypes.  

Convolutional neural networks have been extended to graph-based fusion. Multi-omics graph convolutional network (MORONET) is a multi-omics fusion framework for classification that utilizes supervised convolutional networks for omics datasets.\cite{wang2020moronet} Multi-omics graph convolutional network (M-GCN) is another multi-omics fusion method developed based on convolutional neural networks for molecular subtyping.\cite{yin2022molecular}

Despite the strengths of neural network-based fusion methods, their performance depends on a large training sample size that is of limited availability in the multi-omics field. Moreover, the neural network-based fusion approach lacks interpretability, an essential need for biologists to identify biological functions.\cite{bodein2022interpretation}

\subparagraph {Probabilistic model-based fusion} is commonly based on the hidden Markov model (HMM) to make probabilistic models that encode information as transition matrices and mix them for future tasks. \cite{bayoudh2022survey} Markov chain models variables (a.k.a. states) and transition probabilities between states to produce a sequence of observations. The transition probabilities indicate the probability of moving from one state to another. \cite{NIPS1995_4588e674} The ability of HMM to consider the correlations between states makes it an effective approach for analyzing multi-omics data. \cite{yoon2009hidden} The use of HMMs in biological analysis has been investigated in several works. \cite{gentili2022biological,yoon2009hidden}

Another form of probabilistic model-based fusion is the Bayesian network approach, which constructs a directed acyclic graph to represent the probability distribution of each omics. \cite{subramanian2020multi} \citet{fridley2012ab} used a Bayesian hierarchical structure to fuse multiple types of genomics data with phenomics data to find genomics’ direct and indirect effects on the phenotype. \citet{wang2019bayesian} introduced a Bayesian framework to combine genomics, transcriptomics, and epigenomics data for identifying the high-confidence risk genes of schizophrenia. As another example, \citet{zhang2022genome} developed a machine learning method, regional fine-mapping (RefMap), which is a hierarchical Bayesian framework for gene discovery in amyotrophic lateral sclerosis. In their work, epigenomics and transcriptomics have been integrated for gene discovery. 

\paragraph{Factorization.}
Fusion based on factorization takes multiple omics modalities as input matrices and decomposes them into two parts: (i) factors that are common to all omics and (ii) weights for each modality. Common factors can be utilized for patient clustering, and weights help identify biomarkers.\cite{15_17} The decomposition assumes that biological mechanisms can be detected by biological factors shared among multiple modalities.\cite{picard2021integration,huang2017more} Therefore, this type of fusion is capable of acquiring a complex inter-omics structure. Two approaches designed to perform factorization are covered in the following subsections. 

\subparagraph {Matrix factorization fusion} factorizes multi-omics data matrices into the product of several matrices, including omics-specific weight matrices and a factor matrix.\cite{15_17} As a result, data are projected into a shared latent space to find driving factors for diseases. In unimodal learning, the most popular matrix factorization is PCA, which decomposes the covariance matrix of data to extract underlying biological factors.\cite{bishop2006pattern} Various methods have been developed to generalize PCA for multi-omics fusion. Multi-omics factor analysis (MOFA),\cite{argelaguet2018multi} joint and individual variation explained (JIVE),\cite{lock2013joint} joint non-negative matrix factorization (jNMF),\cite{zhang2012discovery} and integrative non-negative matrix factorization (iNMF)\cite{yang2016non} are generalized PCA-based methods in which multiple omics modalities of the same biological samples are included in the analysis so features in each modality differ. In contrast, multi-study factor analysis (MSFA)\cite{de2019multi} is another generalization wherein the same omics features from different biological samples obtained in multiple studies are included in the analysis. Additionally, several other works have introduced integration methods based on matrix factorization, including iCluster\cite{shen2009integrative} and its extension iCluster+\cite{mo2013pattern} by utilizing maximum likelihood estimation and regularized generalized canonical correlation analysis (RGCCA).\cite{tenenhaus2011regularized}

Although matrix factorization has been extensively investigated in the literature, most existing methods considered a global shared space among omics modalities while neglecting partial common structures; a variable can be shared by two omics modalities but is not available in the third one.\cite{yang2016non,gaynanova2019structural} For example, \citet{gaynanova2019structural} presented structural learning and integrative decomposition method (SLIDE) to model partially shared structures for multi‐view fusion.  

\subparagraph {Tensor-based factorization} typically constructs higher-order relationships among biological variables to extract factors that play essential roles in describing these relationships. \cite{5} In other words, omics modalities are presented in a higher dimensional space in which the new dimension indicates the data modality. \cite{Kolda2009tensor} \citet{jung2021monti} introduced a two-stage tensor-based factorization method (MONTI) for multi-omics analysis in cancer subtyping. In the first stage, non-negative tensor factorization was used to factorize tensors constructed from multi-omics data, and in the second stage, a representative feature subset was selected using L1 regularization. 

Tensor-based factorization fusion can be computationally expensive, especially as the number of omics modalities increases. \cite{kuleshov2015tensor,5} \citet{teschendorff2018tensorial} proposed a tensorial independent component analysis (tICA) based on independent component analysis. They used data tensors of order 4 for the epigenome-wide association studies (EWAS) dataset, which resulted in better efficiency in comparison to the current methods. Moreover, a number of attempts have been made to use Bayesian inference in tensor factorization. \cite{tang2018bayesian,liu2022bayesian}

\paragraph{Joint feature selection.}
\label{sec:fusion-joint}
This approach selects features with joint consideration of multiple omics modalities in the integration process. In most existing multi-omics fusion methods, feature selection is independently applied to each omics modality as a pre-processing step before integration. These methods reduce feature space independently so that the relationships between multiple omics could be lost during such pre-processing.\cite{picard2021integration,tabakhi2022magentomics} 

Several works have explored joint feature selection for multi-omics integration that can be categorized into single-point strategy and multi-agent systems. 

\subparagraph {The single-point strategy} aims to select a subset of features from the whole omics data by starting from a specific point. This strategy iteratively adds new features selected from each omics data according to a statistical metric. \citet{el2018min} presented a joint feature selection model applied to a multi-view cancer dataset. Their main idea was to generalize the minimum-redundancy and maximum-relevance statistical method developed for single-view feature selection to multi-omics research via an incremental process. 

Although this fusion strategy considers the correlation between omics data, its sensitivity to the starting feature is likely to result in limited performance.\cite{tabakhi2015relevance}

\subparagraph {Multi-agent systems} (MAS) are an improvement over the single-point strategy in which several starting points are simultaneously selected to guide the feature selection procedure. In MAS, several independent agents collaborate with each other in a shared environment to solve a complex problem. The distributed and parallel problem-solving abilities, using knowledge of other agents through interactions, decision-making flexibility of individual agents, and reliability are essential features of MAS to handle complex problems. \cite{dorri2018multi}

Recently, \citet{tabakhi2022magentomics} introduced a novel multi-agent system for multi-omics data integration through an unsupervised feature selection approach. This method represented each omics data as a fully connected weighted graph. Then, agents communicated with each other and shared their knowledge to select the best feature subset from all omics datasets using an iterative procedure to improve performance. 

\subsection{Decision-level fusion}
\label{sec:late-fusion}
Fusion at the decision level (also known as late fusion) builds multiple machine learning models independently on each omics modality and then aggregates predictions from these models for the final decision. This approach is flexible because different machine learning models can be constructed for each omics modality.\cite{adossa2021computational} Majority voting, hierarchical classifiers, and ensemble-based methods are the most extensively used aggregators at the level of decision.\cite{carrillo2022machine,sharifi2019moli,huang2019salmon,miao2021multi}

The ability to integrate various single-omics frameworks to build multimodal learning algorithms is the key strength of decision-level fusion.\cite{picard2021integration} In addition, having separate learning models enables this fusion approach to handle heterogeneity across multiple modalities. Because learning algorithms are independently trained on each omics data modality, late fusion can also address the feature imbalance problem.

\section{Open-Source Software}
\label{sec:tools}
Despite a growing number of proposed methods in the literature, only a limited number of researchers have published the source code of their works, particularly in the earlier years. Moreover, because each developer follows different protocols while writing code, others may find it challenging to utilize the available code to conduct fair experiments for their studies. Standardizing codes by publishing research software with the implementation of state-of-the-art methods facilitates the community to reuse. This issue is rarely addressed in existing surveys on multi-omics.

This section presents a collection of open-source software that has implemented several multimodal learning algorithms for multi-omics. This collection has been based on the following criteria:  
\begin{itemlist}
 \item Programming language: This criterion shows the popularity of each programming language among developers in this field. 
 \item Creation and last update: These two metrics indicate the recency and the software age over time. 
 \item GitHub stars: This measure demonstrates the reputation and influence of the software in the community.
 \item Paper citations: This indicator helps understand the prevalence of the software in the current research.
\end{itemlist}

Table \ref{table:tools} summarizes a list of open-source software collected based on the above criteria by explaining covered modalities, used datasets, their functionality, and a link to their code repositories. In this table, the omics/modalities column refers to different types of omics, the datasets column indicates the names of multi-omics datasets used by the software, and the functionality column explains the various usages of the software.

By introducing this collection, we make researchers more aware of existing software and encourage developers to share the software of their methods on open-source platforms. Moreover, we believe that standardizing software in multi-omics research can benefit from the presentation of this list, for example, via integrating existing software into benchmark software.

\afterpage{%
\begin{landscape}
\begin{table}[th]
\vspace*{-1cm}
\small
  \caption{Open-source software tools for multi-omics analysis as of August 2022.}
  \renewcommand{\arraystretch}{1.4}
  \label{table:tools}
  \centering
    \hspace*{-2cm}
    \makebox[\textwidth]{\resizebox{1.7\textwidth}{!}{\begin{tabular}{p{0.1\linewidth}p{0.1\linewidth}p{0.16\linewidth}p{0.22\linewidth}p{0.06\linewidth}p{0.055\linewidth}p{0.06\linewidth}p{0.08\linewidth}p{0.32\linewidth}p{0.25\linewidth}}
    \toprule
    Software tool & Programming language & Omics/Modalities & Datasets  & Creation & Last update & GitHub stars  & Paper \newline citations & Functionality & Repository link \\
    \midrule
    MultiBench \cite{5} & Python & 
    Language \newline
    Image \newline
    Video \newline
    Audio \newline
    Time-series \newline
    Tabular \newline
    Force sensors \newline
    Proprioception sensor \newline
    Optical flow
    & 
    MUStARD \cite{6} \newline
    CMU-MOSI \cite{7}\newline
    UR-FUNNY \cite{8}\newline
    CMU-MOSEI \cite{9}\newline
    MIMIC \cite{10}\newline
    MuJoCo Push \cite{11}\newline
    Vision \& Touch \cite{12}\newline
    Stocks-F\&B\newline
    Stocks-Health\newline
    Stocks-Tech\newline
    ENRICO \cite{13}\newline
    Kinetics400-S \cite{14}\newline
    MM-IMDb \cite{15_17}\newline
    AV-MNIST \cite{16}\newline
    Kinetics400-L \cite{14}
     &2021& 2022 & 181&13 & MultiBench presents a unified pipeline to simplify and standardize data loading, experimental setup, and model evaluation for machine learning processes. MultiBench aims to evaluate domain generalization, time and space efficiency, and model robustness. & \url{https://github.com/pliang279/MultiBench} \\
    Momix\cite{15_17} & R & 
    Transcriptomics\newline
    Genomics
    & 
    TCGA \cite{18} \newline
    Single-cell data \cite{19} 
    &2019& 2022 &48&18& Momix has implemented several representative joint dimensionality reduction methods for multi-omics analysis. This tool focuses on extracting biological information, clustering samples, finding pathways/biological functions, and extracting representative features.& \url{https://github.com/cantinilab/momix-notebook} 
    \\
    Mergeomics \cite{53}& R&Genomics\newline
    Transcriptomics\newline
    Proteomics\newline
    Metabolomics&GWAS \cite{54}\newline
    TWAS \cite{li2021gwas}\newline
    EWAS \cite{li2021gwas}
    &2019&2022&3&33& The Mergeomics package has been developed with the aim of integrating multidimensional data to uncover disease-associated pathways and networks through a flexible pipeline.&\url{https://github.com/jessicading/mergeomics }
    \\
    MiBiOmics \cite{20} & R & 
    Taxonomics\newline
    Proteomics
    &TCGA \cite{18}\newline
    TOE \cite{21}
    &2019& 2022& N/A&15&MiBiOmics is a standalone and web-based application to analyze single or multi-omics data. The application’s pipeline enables multi-omics data exploration, integration, and analysis for broad users.& \url{https://gitlab.univ-nantes.fr/combi-ls2n/mibiomics }
    \\
    mixOmics \cite{46} & R& Transcriptomics\newline
    Metabolomics\newline
    Metagenomics\newline
    Proteomics& TCGA \cite{18}\newline
    Yeast \cite{29}&2018&2022&106&1361&mixOmics has presented the implementation of many multivariate methods for the exploration and integration of biological data, as well as dimensionality reduction and visualization.&\url{https://github.com/mixOmicsTeam/mixOmics }
    \\
    OmicsPLS \cite{47}& R & Transcriptomics\newline
    Metabolomics\newline
& CROATIA Korcula cohort \cite{48}&2018&2022&24&26&OmicsPLS is a free, open-source implementation of the O2PLS algorithm for heterogeneous data integration that is capable of handling both low and high-dimensional datasets.&\url{https://github.com/selbouhaddani/OmicsPLS }
    \\
  \end{tabular}}}
\end{table}
\end{landscape}
\clearpage   %
}

\afterpage{%
\begin{landscape}
\begin{table}[th]
\ContinuedFloat
\vspace*{-1cm}
\small
\caption{Open-source software tools for multi-omics analysis as of August 2022. (Continued)}
\renewcommand{\arraystretch}{1.4}
  \centering
    \hspace*{-2cm}
    \makebox[\textwidth]{\resizebox{1.7\textwidth}{!}{\begin{tabular}{p{0.115\linewidth}p{0.1\linewidth}p{0.16\linewidth}p{0.22\linewidth}p{0.06\linewidth}p{0.055\linewidth}p{0.06\linewidth}p{0.08\linewidth}p{0.32\linewidth}p{0.25\linewidth}}
    \toprule
    Software tool & Programming language & Omics/Modalities & Datasets  & Creation & Last update & GitHub stars  & Paper \newline citations & Functionality & Repository link \\
    \midrule
    Ingenuity Pathway Analysis \cite{32} & R & 
    Transcriptomics\newline
    Metabolomics\newline
    Proteomics\newline
    Metagenomics & 
    GSE11352 \cite{33}\newline
    GSE2639 \cite{34}&2014&2022&N/A&2900+&Ingenuity Pathways Analysis (IPA) aims to carry out intelligent multi-omics data analysis and interpretation in the context of biological systems.& \url{https://analysis.ingenuity.com/pa/installer/select }
    \\
    COBRA \cite{22}& Matlab & 
    Proteomics\newline
    Transcriptomics\newline
    Fluxomics\newline
    Metabolomics
    &Human Protein Atlas \cite{23} \newline
    Drugbank \cite{24} 
    &2012& 2022 & 196&370&COBRA software has developed a collection of advanced methods for the integrative analysis of molecular data, specifically the genotype-phenotype relationship. COBRA protocols allow it to be used on any biomedical system. & \url{https://github.com/opencobra/cobratoolbox/ }
    \\
    MetaboAnalyst \cite{44} & R & 
    Transcriptomics\newline
    Genomics\newline
    Proteomics\newline
    Metabolomics
    &HMDB \cite{45}&2011&2022&207&2500+&MetaboAnalyst is a web-based tool that analyzes metabolomics data and its correlations with other omics data statistically and functionally.&\url{https://github.com/xia-lab/MetaboAnalystR}
    \\
    IMPaLA \cite{30} & Python & 
    Proteomics\newline
    Transcriptomics\newline
    Metabolomics & 
    NCI60 \cite{31} &2011& 2021 & N/A&264& IMPaLA is a web tool for the joint pathway analysis of multiple omics types, including transcriptomics, proteomics, and metabolomics. This tool provides its functionality through a SOAP web service as well. & \url{http://impala.molgen.mpg.de/ }
    \\
    E-Cell \cite{25} & C++\newline Python & 
    Proteomics\newline
    Genomics\newline
    Metabolomics
    & Mycoplasma genitalium gene set \cite{26}
    &2010& 2021 & 61&593 & E-Cell is a platform that enables modeling, simulating, and analyzing complex multi-scale systems. & \url{https://github.com/ecell/ecell4_base}
    \\
    iDINGO \cite{57} &R & 
    Transcriptomics\newline
    Metabolomics\newline
    Metagenomics\newline
    Proteomics&
    TCGA \cite{18}&2017&2020&4&9&iDINGO is a package developed for group-specific conditional dependencies estimation and differential network analysis between groups. This analysis is carried out within a single or integrative framework.&\url{https://github.com/cran/iDINGO  }
    \\
    Escher \cite{58} & JavaScript\newline
    Python\newline
    Java\newline & 
    Genomics\newline
    Metabolomics\newline
    Proteomics\newline
    Transcriptomics\newline
    Fluxomics & N/A  &2013& 2019 & 176&231&Escher is a web-based platform designed for creating new pathways quickly based on pathway suggestions and visualizing biological pathways (i.e., reactions, genes, and metabolites). & \url{https://github.com/zakandrewking/escher } 
    \\
    Recon3D \cite{49}&Python \newline Matlab&
    Proteomics\newline
    Genomics\newline
    Metabolomics&
    HMR 2.0 \cite{50}\newline
    Protein Data Bank(PDB) \cite{51}\newline
    CHEBI \cite{52}&2017&2018&18&299&Recon3D has been developed for the integrative analysis of pharmacogenomic associations, large-scale phenotypic data, and structural information for proteins and metabolites. It allows the prioritization of genetic variations that cause disease.&\url{https://github.com/SBRG/Recon3D}
    \\
    GIM3E \cite{27}& Python & 
    Metabolomics \newline
    Transcriptomics & 
    KEEG\cite{38} \cite{28}\newline
    Yeast \cite{29} &2013& 2016 &3&99&GIM3E is a tool for implementing the integrative functional analysis of metabolomics and gene expression microarray data. &
    \url{https://github.com/brianjamesschmidt/gim3e }
    
  \end{tabular}}}
\end{table}
\end{landscape}
\clearpage   %
}

\afterpage{%
\begin{landscape}
\begin{table}[th]

\ContinuedFloat
\vspace*{-1cm}
\small
\caption{Open-source software tools for multi-omics analysis as of August 2022. (Continued)}
\renewcommand{\arraystretch}{1.4}
  \centering
    \hspace*{-2cm}
    \makebox[\textwidth]{\resizebox{1.7\textwidth}{!}{\begin{tabular}{p{0.1\linewidth}p{0.1\linewidth}p{0.16\linewidth}p{0.22\linewidth}p{0.06\linewidth}p{0.055\linewidth}p{0.06\linewidth}p{0.08\linewidth}p{0.32\linewidth}p{0.25\linewidth}}
    \toprule
    Software tool & Programming language & Omics/Modalities & Datasets  & Creation & Last update & GitHub stars  & Paper \newline citations & Functionality & Repository link \\
    \midrule

    3Omics \cite{37} & Perl \newline PHP & 
    Transcriptomics \newline
    Metabolomics \newline
    Proteomics & 
    KEGG \cite{38} \newline
    HumanCyc \cite{39}\newline
    DAVID \cite{40}\newline
    Entrez Gen3 \cite{41}\newline
    OMIM and UniProt \cite{42}&2012& 2013& N/A&112&3Omics is a web-based application for analyzing and visualizing multiple omics data, including transcriptomics, proteomics, and metabolomics. This application consists of several widely used analyses such as correlation network, co-expression, phenotype generation, KEGG/HumanCyc pathway enrichment, and GO enrichment. & \url{https://3omics.cmdm.tw}
    \\
    MapMan \cite{35} & R & 
    Transcriptomics\newline
    Metabolomics\newline
    Metagenomics
    & TAIR \cite{36}\newline
    KEGG \cite{38}&2010& 2013&N/A&518&MapMan provides a platform for classifying genes and metabolites and visualizing the results in relation to pathways and processes.& \url{https://mapman.gabipd.org/mapman }
    \\
    Cell Illustrator \cite{43} & Java & 
    Transcriptomics \newline
    Genomics\newline
    Proteomics & N/A&2006&2010&N/A&69&Cell Illustrator tool allows biologists and biochemists to use the concept of Petri net for modeling (metabolic pathways and gene regulatory pathways), visualizing biological processes, and interpreting the results.& \url{https://cionline.hgc.jp/cifileserver/apps/usersman/main}
    \\
    ImmuNet \cite{55} & R&
    Proteomics\newline
    Transcriptomics\newline
    Metabolomics&
    KEGG \cite{38}\newline
    BioGRID \cite{56}&2015& N/A &N/A&40&ImmuNet provides an interactive and configurable web-based platform for researchers to easily explore numerous functional relationship networks related to the immune system.&\url{http://immunet.princeton.edu/ }
    \\
    \bottomrule
  \end{tabular}}}
\end{table}
\end{landscape}
\clearpage   %
}
 
\section{Multi-Omics Datasets}
\label{sec:dataset}
The accessibility of multi-omics datasets is vital for researchers to work on omics data challenges and obtain a deeper insight into diseases. One crucial step in conducting fair comparisons in the literature is utilizing public multi-omics datasets. With the growing number of datasets in the research field, it is helpful to curate a list of well-known and frequently used datasets. Moreover, with the variety of datasets, standardizing them on a platform can support the community to search and download datasets easily. We summarize these findings in the following two subsections.  

\subsection{Datasets description}
In this section, we briefly review important multi-omics datasets such as TCGA, \cite{78,79,80} Kyoto Encyclopedia of Genes and Genomes (KEGG), \cite{38,85,86} and the Human Protein Atlas (HPA). \cite{uhlen2015tissue}

TCGA was initiated by the effort between the National Cancer Institute (NCI) and the Center for Cancer Genomics Research Institute (NHGRI) with the aim of collecting, molecularly characterizing, and analyzing many cancers in 2006. This initiative has processed more than 20,000 patients to represent 33 cancer types and provided over 2.5 petabytes of data from different modalities, including genomics, epigenomics, transcriptomics, and proteomics. According to Google Scholar metrics, TCGA has been cited over 38,500 times as of July 2022, demonstrating the project's popularity among researchers. Another initiative is the KEGG program, conducted at Kyoto University as part of the Human Genome Program in 1995. The program's objective is to assign sets of genes in the genome with higher-order functional information that can help the biological interpretation of genomic information. KEGG has analyzed different omics types, including genomics, transcriptomics, proteomics, metabolomics, and other types. As another example, HPA collects transcriptomics and proteomics data for human tissue expression profiles. Table \ref{table:dataset} lists other studies that focus on providing multi-omics data to enable the development of multi-omics fusion workflows.

\subsection{Online portals/platforms}
In addition to the software outlined in Section \ref{sec:tools} for facilitating the integration of multi-omics datasets, numerous portals/platforms exist for downloading multiple omics modalities. These portals allow researchers to explore and access various omics modalities easily and optimize the reuse of datasets. Table \ref{table:platforms} summarizes popular online portals in multi-omics research that can help the community discover datasets for their research and development.\\

\afterpage{%
\begin{table}[th]

\small
\caption{Multi-omics benchmark datasets collected by August 2022.}
\renewcommand{\arraystretch}{1.4}
\hspace*{-1.5cm}
\label{table:dataset}
{\begin{tabular}{@{}p{0.23\linewidth}p{0.2\linewidth}p{0.25\linewidth}p{0.1\linewidth}p{0.37\linewidth}@{}} \toprule
Dataset & Omics type & Target & Citations & Web link 
\\ \colrule
TCGA\cite{78,79,80,81,82,83} & Genomics\newline Epigenomics\newline Transcriptomics\newline Proteomics & 33 types of cancer & 38532& \url{https://portal.gdc.cancer.gov/}
\\
KEGG\cite{38,85,86} & Genomics\newline Metabolomics & Infectious diseases & 25554& \url{https://www.genome.jp/kegg}
\\
HPA\cite{uhlen2015tissue}& Genomics\newline Transcriptomics\newline Proteomics & More than 17 types of cancer& 8166& \url{https://www.proteinatlas.org/}
\\
HMDB\cite{89} & Metabolomics\newline Proteomics & Kidney disease & 7471& \url{https://hmdb.ca/}
\\
PDB\cite{90,91,92} & Proteomics\newline Genomics & Topical health matters (e.g., Zika, measles, coronavirus) & 4044& \url{https://pdb101.rcsb.org/}
\\
NSCLC\cite{93,94} & Genomics\newline Radiomics & Lung cancer& 3551& \url{https://wiki.cancerimagingarchive.net/display/Public/NSCLC-Radiomics-Genomics} \\

NCI-60\cite{101} & Genomics\newline Transcriptomics & 60 types of human tumors and cancers & 2700& \url{https://dtp.cancer.gov/discovery_development/nci-60/} \\

ICGC\cite{102} &Genomics\newline Transcriptomics & Lung, gynecologic, breast, gastrointestinal, squamous, and renal cancers& 2039& \url{https://dcc.icgc.org/} \\

YeastMine & Genomics\newline Transcriptomics & Oxidative stress and meningitis & 284& \url{https://yeastmine.yeastgenome.org/yeastmine/begin.do} \\
Mycoplasma genitalium\cite{99} & Genomics\newline Proteomics & Pelvic Inflammatory Disease (PID) & 200& \url{https://www.ncbi.nlm.nih.gov/Taxonomy/Browser/wwwtax.cgi?mode=Info\&id=txid2097}
\\
GSE11352\cite{33} & Transcriptomics\newline Genomics & Breast cancer & 177& \url{https://www.ncbi.nlm.nih.gov/geo/query/acc.cgi?acc=GSE11352} 
\\
GWAS\cite{97} & Genomics\newline Transcriptomics\newline Proteomics\newline Metabolomics\newline Epigenomics & Diabetes, heart diseases, Parkinson, and Crohn & 53& \url{https://ngdc.cncb.ac.cn/gwas/} 
\\
 \botrule
\end{tabular}}
\end{table}
\clearpage   %
}

\afterpage{%
\begin{landscape}
\begin{table}[th]
\vspace*{-1cm}
\small
\caption{Popular online portals/platforms for multi-omics analysis.}
\renewcommand{\arraystretch}{1.4}
\label{table:platforms}
  \centering
    \hspace*{-2cm}
    \makebox[\textwidth]{\resizebox{1.7\textwidth}{!}{\begin{tabular}{@{}p{0.2\linewidth}p{0.5\linewidth}p{0.25\linewidth}p{0.2\linewidth}@{}}
    \toprule
    Platform name & Description & Data &
Web link \\
    \midrule
    UCSC Xena \cite{goldman2020visualizing}&Provide users with an interactive online portal for the exploration of cancer genomics datasets &Over 1,600 datasets from over 50 cancer types &\url{https://xena.ucsc.edu/}
\\
EMBL-EBI OLS \cite{Jupp2015ANO} & Allow users to search, browse, and visualize biomedical ontologies based on their meta-data from a centralized location & Biomedical ontologies& \url{https://www.ebi.ac.uk/ols/index}
\\
OmicsDI\cite{perez2017discovering} &Present an open-source portal for accessing, discovering, and disseminating omics datasets & Over 100,000 datasets from 16 different public data resources&\url{https://www.omicsdi.org/}
\\
ACGT\cite{rappoport2018multi} & Offer several preprocessed TCGA multi-omics data benchmarks & Several TCGA multi-omics data & \url{http://acgt.cs.tau.ac.il/multi_omic_benchmark/download.html}
\\
ICGC Data Portal\cite{zhang2019international} & Provide an interactive web-based platform to store, annotate, and explore large and complex cancer genomics datasets& Over 84 worldwide cancer projects&\url{https://dcc.icgc.org/repositories }
\\
GDC Data Portal \cite{grossman2016toward}&Enable users to efficiently query and download high-quality and complete cancer data for analysis&Genomic and clinical data from cancer research programs (e.g., TCGA, TARGET, CGCI)&\url{https://portal.gdc.cancer.gov/repository }
\\
The Cancer Imaging Archive\cite{clark2013cancer}&Present an open-source, open-access service allowing public access to a large available archive of cancer-related medical images&Archive of medical images of cancer&\url{https://www.cancerimagingarchive.net/collections/ }
\\
JGI Genome Portal\cite{nordberg2014genome}& Provide a unified platform to search, download and explore JGI genomic databases& JGI genomic databases & \url{https://genome.jgi.doe.gov/portal/ }
\\
European Genome-Phenome Archive\cite{freeberg2022european}&Offer a platform for storing and sharing genetic, phenotypic, and clinical data generated for biomedical research projects& The repository of over 6,800 human datasets & \url{https://ega-archive.org/}
\\
Depmap Portal\cite{tsherniak2017defining}&Allow the community to explore genetic and pharmacological dependencies&CRISPR knockout screens from project Achilles and genomic data from the CCLE project&\url{https://depmap.org/portal/download/ }
\\
Canadian VirusSeq Data Portal &Provide an open-source platform for all SARS-CoV-2 sequences and associated non-personal contextual data collected in Canada&SARS-CoV-2 genome sequences among Canadian &\url{https://virusseq-dataportal.ca/explorer }
\\
iLINCS\cite{pilarczyk2022connecting}&Present a web-based portal for exploring an extensive collection of omics datasets (i.e., transcriptomics and proteomics) and cellular perturbation signatures& More than 34,000 processed omics datasets and more than 220,000 omics signatures&\url{http://www.ilincs.org/ilincs/datasets/portals }
\\
LinkedOmics\cite{vasaikar2018linkedomics}&Present a web tool for biologists and clinicians to access, analyze, and compare multi-omics data within and across tumor types&32 TCGA multi-omics data and 10 CPTAC cancer cohorts&\url{http://www.linkedomics.org/ }
\\
\botrule
  \end{tabular}}}
\end{table}
\end{landscape}
\clearpage   %
}

\section{Discussion and Future Directions}
\label{sec:discussion}
Multimodal learning for multi-omics has become increasingly attractive in biomedical studies, especially with the availability and easy access to multiple omics modalities.\cite{gligorijevic2015methods} Multi-omics analysis offers growing opportunities for answering key biological questions by providing deep views on the underlying mechanisms of diseases. Therefore, extensive methods have been proposed to analyze the multiple biomedical data sources. The community needs to be aware of the recent methodologies for mitigating multi-omics data challenges, analytical methods, open-source software, and benchmark datasets to conduct comprehensive and reproducible experiments. Several recent attempts have been made to survey the field of multi-omics from different points of view, as summarized in Table \ref{table:survey}. However, as shown in Table \ref{table:survey}, a comprehensive review is still needed to provide an organized overview of current data challenges, introduce a more structured taxonomy of multi-omics approaches to cover recent works, present well-known and widely used datasets in the field with the available platforms to access them, and collect active software for analyzing multi-omics data. In this study, we have made efforts to address these needs, and the following interesting points deserve attention to progress in the research toward a better future.

\begin{itemlist}
\item \textbf{Integrate more omics modalities}: Multi-omics data consist of molecular, imaging, and phenotypic data. Molecular data capture the information of different molecular layers, and many works in the literature have been carried out to fuse multiple molecular omics modalities. On the other hand, medical images have emerged from imaging techniques to strengthen the multi-omics research field by presenting meaningful information on disease diagnosis and treatment. Several frameworks have been widely developed to fuse medical images with gene expression data, creating a new field of study termed `radiogenomics.'\cite{lo2020combining,bodalal2019radiogenomics} Another source of omics data is phenomics, which is information about the phenotypic characteristics of individuals. This valuable dimension of data has been chiefly integrated into genome data, resulting in the creation of an informatics research direction named `Phenoinformatics.'\cite{xu2015plant}

Since the development of human diseases involves complex interactions of biological mechanisms, a comprehensive snapshot of the diseases can be obtained by analyzing different levels of information. A critical question in analyzing multi-omics data is whether adding new omics data can boost our ability to perform biological tasks. However, with the multiple sources of omics modalities, researchers have been exploring integrating two dimensions of omics data simultaneously  (bimodal fusion\cite{acosta2022multimodal}), as reviewed in Section \ref{sec:fusion}. Therefore, we encourage the community to generalize their methods across all omics modalities to benefit from their complementary information.

\item \textbf{Find better solutions for data challenges}:  Figure \ref{fig:challenges} has summarized many multi-omics data challenges, which have been the focus of many studies to facilitate further analysis. The curse of dimensionality is a critical challenge in multi-omics analysis that must be overcome. The presence of irrelevant and redundant features in multi-omics data makes it challenging to acquire informative patterns.\cite{hira2015review,kirpich2018variable} Many methods have separately applied dimensionality reduction techniques to each omics modality without taking inter-omics connections into account. As pointed out in Section \ref{sec:fusion-joint}, the utilization of feature selection on all omics modalities has been performed in several works,\cite{el2018min,tabakhi2022magentomics} which are computationally expensive for large-scale data. \textcolor{red}{Moreover, leveraging graph representation learning to map input omics data into a low-dimensional space enables automatic dimensionality reduction, which can be helpful for downstream tasks.} Therefore, proposing new efficient methods to consider interactions between multiple omics in an affordable computational efficiency will benefit such research. 

Another challenge is class/feature imbalance. Imbalance at the class level is a traditional machine learning challenge, and many approaches have been proposed to address it, such as over and under-sampling and cost-sensitive learning.\cite{lin2016imbalanced,wang2022large,johnson2019survey,zhou2005training} At the level of features, the imbalance problem specifically occurs in multimodal learning when different modalities with different feature dimensionality are fused. While the feature imbalance challenge can be implicitly dealt with through the late fusion approach, proposing explicit pre-processing strategies will also be helpful. The investigation of dimensionality reduction techniques is another potential direction for solutions to the imbalance challenge.

In the case of missing values, significant efforts are needed for partial datasets. Converting this type of data to a full dataset by removing samples with missing omics is a naive way yet regularly used in studies. The removed samples could contain useful information, which may enhance the outcomes if included. Another interesting area of research is about filling missing values in samples by independent imputation on each omics data.\cite{theodoridis2006pattern,chen2019sparse,feng2021addressing} Integrative imputation can potentially tackle this problem by leveraging the strengths of multiple omics.  

Another specific characteristic of multi-omics data is heterogeneity. Various data distributions and different data types are found in the heterogeneous data. Some work has been introduced to normalize and scale the omics data properly.\cite{krassowski2020state,tarazona2020harmonization,Akdemir857425} Further development to deal with this challenge can be achieved by proposing accurate and appropriate fusion approaches. %

Furthermore, multi-omics data are usually noisy and can often have mislabeled samples. Even though several error correction approaches have been successfully developed,\cite{yoo2021community,huang2013tool,lee2019probabilistic,javed2020detecting} leveraging unsupervised/self-supervised learning could shed new light on tackling this challenge.

\item \textbf{Investigate new fusion approaches}: From a multi-omics data fusion point of view, there are a vast number of methods for integrating multi-omics data. Some of these methods have been generalized from other domains to multi-omics analysis. Although they have strongly performed in their domains, omics-specific properties can impose additional challenges.\cite{5} There have been several attempts to build hierarchical models to incorporate the sequential organization among omics.\cite{fortelny2020knowledge,browaeys2020nichenet,weidemuller2021transcription} These models have been constructed based on the cause-effect relationships discovered in the molecular data. However, they rely on prior knowledge of omics interactions provided by experts,\cite{picard2021integration} which is challenging but deserves careful investigation in the future.

On the other hand, a new category of fusion approaches is joint feature selection, as discussed in Section \ref{sec:fusion-joint}. This category is a promising approach for combining omics data through a multimodal feature selection scheme. Despite its superior performance in diagnostic tasks, this approach requires new attempts to boost computational efficiency to deal with large-scale datasets.

Although it is important to integrate new omics data, several challenges can arise more severely, such as higher dimensionality of features, more noise for the learning model, and lower efficiency. To tackle these issues, new fusion methods need to be scalable with respect to the size of omics data. Based on the review in Section \ref{sec:fusion-mix}, the transformation-based approach, particularly graph representation learning, can be scalable and is among the promising ways to develop new fusion methods.

\item \textbf{Develop open-source multi-omics analysis software}: Recently, open-source software has been increasingly used to develop complex real-world systems. At the same time, the area of machine learning has created a large number of efficient algorithms for a variety of applications. After analysis, we found that one major gap in Table \ref{table:tools} on open-sourced multi-omics data integration software is that most do not have up-to-date codebases or documentation. 
One reason was that there were few open-sourced software platforms earlier, and they did not have version control like GitHub. An open-source codebase and well-documented software are critical for \textcolor{blue}{further extension} or modification. Therefore, to grow the field of integrative multi-omics analysis, we need more open-source software with high-quality documentation to enable and encourage more researchers to contribute to this field, thus lowering the entry barriers.

As explained earlier, one major future research direction that can add great value to multi-omics data analysis is the incorporation of radiomics and phenomics data. Limited software is available for fusing radiomics and phenomics with other molecular omics data. It can be seen from Table \ref{table:tools} that no software listed has used radiomics or phenomics data. The interoperability of various omics datasets needs to be enhanced to make them more accessible. A unified software framework will help with the seamless integration of such biomedical data.

\end{itemlist}

\textcolor{red}{To further advance multi-omics research, we highlight more succinctly the following key insights based on the current progress of the field to fill the gaps promptly and efficiently in future works: 
\begin{romanlist}[(ii)]
 \item Integrating multiple modalities, including molecular, imaging, and phenotypic data, will drive the progress of the multi-omics studies forward by capturing their complementary information. Community-based standardization is a key enabler for such integration. For example, the European life science infrastructure for biological information (ELIXIR)\cite{crosswell2012elixir} was initiated with the aim of providing a sustainable infrastructure for standardizing data to facilitate easy and open access to biological data. Because data availability boosts the strengths of multimodal analysis, we encourage researchers to contribute to these communities to accelerate open and reproducible research.
 \item Even with data standardized, great efforts are needed to mitigate the multi-omics-related data challenges. For the curse of dimensionality, it is crucial to consider intra-omics and cross-omics interactions between different modalities. Modeling the interactions between omics modalities by leveraging graph embedding models or joint feature selection schemes is an exciting study area. In addition, we expect that the integrative imputation achieved by taking multiple omics into account will improve the machine learning model's capability to handle missing values. Finally, using unsupervised/self-supervised learning to overcome data noise will enable effective multi-omics analysis.
 \item With data challenges adequately handled, the next focus is the fusion of different omics modalities. Promising strategies include constructing hierarchical models through cause-effect relationships, leveraging graph embedding models, and jointly selecting a subset of features. Importantly, novel fusion models should be biologically interpretable and robust since they are at least as significant as the model's accuracy.
 \item We finally believe that the development of open-source and well-documented software will promote the growth of the integrative multi-omics analysis field. Several journals, such as Nature Machine Intelligence, enforce the publication of high-quality machine learning open-source software towards standardizing research software that follows the FAIR4RS (findable, accessible, interoperable and reusable principles for research software) guiding principles\cite{chue_hong_neil_p_2022_6623556} that can improve code reusability significantly.
\end{romanlist}}

\section{Conclusion}
\label{sec:conclusion}
With the availability of multiple omics data and their importance to biologists and data scientists in uncovering mechanisms underlying biological conditions, we provided a comprehensive review of multimodal learning for multi-omics from various perspectives: multi-omics data and their challenges, fusion approaches, public datasets and platforms, and open-source software. A brief description of different sources of omics data was presented with real biomedical applications. Moreover, the challenges in multi-omics data were highlighted in detail, and possible solutions to deal with them were introduced. A more structured taxonomy of multi-omics fusion approaches has also been proposed to better categorize currently developed methods. Furthermore, trustworthy and popular multi-omics benchmarks and platforms to download them were introduced. Another dimension we reviewed was active open-source software to help the community move forward faster. Finally, further directions for researchers were discussed to share promising research ideas. 

We hope this survey will help researchers understand better the state of the field in developing efficient and effective solutions to overcome challenges and push forward the study of multi-omics data analysis.

\markboth{S. Tabakhi, M.N.I. Suvon, P. Ahadian, \& H. Lu}{Multimodal Learning for Multi-Omics: A Survey}
\bibliographystyle{IEEEtranN}
\bibliography{ws-wsarai}

\end{document}